\documentstyle[twoside,fleqn,espcrc2]{article}

\newcommand{\AmS}{{\protect\the\textfont2
  A\kern-.1667em\lower.5ex\hbox{M}\kern-.125emS}}

\title{Perturbative Subtraction Methods}

\author{Walter Wilcox\address{Department of Physics, Baylor 
								University, Waco, TX 76798-7316}}
       
\begin{document}

\begin{abstract}
The effects of an automated, tenth order in $\kappa$
subtraction scheme on the noise variance of
various Wilson QCD disconnected matrix elements are examined.
It is found that there is a dramatic reduction in the variance of
the lattice point-split electromagnetic currents
and that this reduction persists at small quark mass.

\end{abstract}

\maketitle

\section{Introduction}

Many types of QCD matrix calculations require
the global estimate of quark \lq\lq disconnected" 
amplitudes. Examples for the nucleon include 
the pi-N sigma term, quark spin content and
the strangeness form factors. There is a continuing 
need for new methods which can extract these matrix elements 
more efficiently.

Noise theory methods are based upon projection of the
signal using random noise vectors as input. That is, given
\begin{eqnarray}
Mx = \eta,
\end{eqnarray}
where $M$ is the $N\times N$ quark matrix, $x$ is the 
solution vector and 
$\eta$ is the noise vector, with
\begin{eqnarray}
<\eta_{i}>=0, <\eta_{i}\eta_{j}>=\delta_{ij},
\end{eqnarray}
where one is averaging over the noise vectors, any 
inverse matrix element,
$M^{-1}_{ij}$, can then be obtained from
\begin{eqnarray}
<\eta_{j}x_{i}>= \sum_{k}M^{-1}_{ik}<\eta_{j}\eta_{k}>=M^{-1}_{ij}.
\end{eqnarray} 

Subtraction methods can be of great assistance in reducing the
noise variance\cite{Thron}. The key is using a 
perturbative expansion of the
quark matrix as the subtraction matrices. The method 
described is completely iterative so that going to higher 
orders in perturbative subtraction is very easy and the
overhead is also extremely low.

\section{The Method}

Let us review the results of noise theory for
the expectation value and
variance of matrices with various types of
noises. Define\cite{Thron}
\begin{equation}
X_{mn} \equiv \frac{1}{L} \sum_{l=1}^{L}\eta_{ml}\eta^{*}_{nl}.
\label{element}
\end{equation}
($m,n=1,\dots ,N$; $l=1,\dots ,L$.) We have
\begin{equation}
X_{mn} =X_{nm}^{*},
\end{equation}
and the expectation value,
\begin{equation}
<X_{mn}>=\delta_{mn}.
\end{equation}
The expectation value of $Tr\{QX\}$ is $Tr\{Q\}$ and it's 
noise variance is
\begin{eqnarray}
V[Tr\{QX\}] \equiv <|\sum_{m,n}q_{mn}X_{nm}-Tr\{Q \}|^{2}>.
\end{eqnarray}
This results in
\begin{eqnarray}
\lefteqn{V[Tr\{QX\}] = \sum_{m\ne n}( <|X_{nm}|^{2}>
|q_{mn}|^{2} + } \\ \nonumber
\lefteqn{<(X_{mn})^{2}> q_{mn}q_{nm}^{*}) +
\sum_{n}<|X_{nn}-1|^{2}>|q_{nn}|^{2}.}
\end{eqnarray}
Consider three cases. First, general real noise: 
\begin{equation}
<|X_{mn}|^{2}>=<(X_{mn})^{2}>=\frac{1}{L},
\label{l}
\end{equation}
for $m\ne n$ so that 
\begin{eqnarray}
V[Tr\{QX_{{\rm real}}\}] = \frac{1}{L}\sum_{m\ne n} 
(|q_{mn}|^{2}
+ q_{mn}q_{nm}^{*}) \\ \nonumber
+\sum_{n} <|X_{nn}-1|^{2}>|q_{nn}|^{2}.
\end{eqnarray}
Compare this with the $Z(2)$ case which also has
Eq.(\ref{l}) for $m\ne n$, but also
\begin{eqnarray}
<|X_{nn}-1|^{2}>=0.
\end{eqnarray}
This shows that
\begin{eqnarray}
V [Tr\{QX_{Z(2)}\}] \le V [Tr\{QX_{{\rm real}}\}]. 
\end{eqnarray}

For the $Z(N)$ ($N\ge 3$) case one has instead
\begin{eqnarray}
<|X_{mn}|^{2}>=\frac{1}{L},\\
<(X_{mn})^{2}>=0,
\end{eqnarray}
for $m\ne n$, but again
\begin{equation}
<|X_{nn}-1|^{2}>=0.
\end{equation}
Thus
\begin{eqnarray}
V[Tr\{QX_{Z(N)}\}] = \frac{1}{L}\sum_{m\ne n} |q_{mn}|^{2},
\end{eqnarray}
and the variance relationship of $Z(2)$ and $Z(N)$ is not fixed in
general for arbitrary $Q$. However, if the phases of $q_{mn}$ and 
$q_{nm}^{*}$ are uncorrelated, 
then $V[Tr\{QX_{Z(2)}\}]\approx V [Tr\{QX_{Z(N)}\}]$ ($N\ge 3$),
which is apparently the case for the operators studied here
and in Ref.~\cite{me}.

Now consider ${\tilde Q}$ such that
\begin{equation}
<Tr\{{\tilde Q} \}>=0.
\end{equation}
Obviously,
\begin{equation}
<Tr\{(Q-{\tilde Q})X \}>=<Tr\{ Q \}>.
\end{equation}
However,
\begin{equation}
V[Tr\{(Q-{\tilde Q})X \}]\ne V[Tr\{Q X \}].
\end{equation}
As we have seen for $Z(N)$ ($N\ge 2$), the variance 
originates exclusively from off
diagonal entries. So the trick is to try to find 
matrices ${\tilde Q}$ which are
traceless (or can be made so) 
but which mimick the
off-diagonal part of $Q$ as much as possible.

The natural choice is simply to choose as ${\tilde Q}$ the 
{\it perturbative}
expansion of the quark matrix. This is given 
by ($\{IJ\}$ are collective indices)
\begin{equation}
(M^{-1}_{p})_{\{IJ\}}=\frac{1}{\delta_{\{IJ\}}-\kappa P_{\{IJ\}}},
\end{equation}
where
\begin{eqnarray}
P_{\{IJ\}}=\sum_{\mu}[(1+\gamma_{\mu})U_{\mu}(x)
\delta_{x,y-a_{\mu}}+ \\ \nonumber
(1-\gamma_{\mu})U_{\mu}^{\dagger}(x-a_{\mu})
\delta_{x,y+a_{\mu}}].
\end{eqnarray}
Expanding this in $\kappa$ gives,
\begin{eqnarray}
\lefteqn{M^{-1}_{p} =  \delta \,\, + \,\,\kappa P  \,\,+\,\, 
\kappa^{2} P^{2}\,\, +\,\,\kappa^{3} P^{3} + 
\cdots ,}\label{one} 
\end{eqnarray}

One constructs $<\eta_{j}
(M^{-1}_{p})_{ik}\eta_{k}>$ and subtracts it from $<\eta_{j}
M^{-1}_{ik}\eta_{k}>$, where
$\eta$ is the noise vector. This construction is
an iterative process and so is easy to code and 
extend to higher powers on the
computer. One can put coefficients in front of the various 
terms in Eq.~(\ref{one}) and vary them to find
the minimum in the variance, but such coefficients take 
on their perturbative
value\cite{Liu} except perhaps for low order
expansions\cite{Dong}. Interestingly, significant variance
improvement occurs in some operators even at 0th order
in $\kappa$.

For a given operator,
${\cal O}$, the matrix ${\cal O}M^{-1}_{p}$ 
encountered in $<{\bar \psi}{\cal
O}\psi>_{gauge}=Tr({\cal O}M^{-1}_{p})$ is not traceless
in general. To correct for this one must re-add
the perturbative part, subtracted earlier, 
to get the full, statistically unbiased answer.
How does one calculate the perturbative part? 
{\it Hard, exact way}:
explicitly construct all the gauge invariant paths 
(up to a given $\kappa$
order) for a given operator. 
{\it Easy, statistical way}: subject the perturbative
contribution to a separate Monte Carlo estimation, 
identical to the Monte Carlo
applied to the nonpertubative part. This separate 
Monte Carlo is easy to do
because one is simply 
constructing a matrix rather than
inverting one. Local operators require perturbative
corrections starting at 4th order and point-split 
ones have corrections
starting at 3rd order. In the following
I will carry this procedure to
$\kappa^{10}$.

\section{Variance Ratio Results}

I will show the ratio of unsubtracted noise variance 
to subtracted variance, $\bf V_{\rm unsub}/V_{\rm sub}$
for $Z(2)$ noise. Since computer time is 
proportional to the operator variance,
the ratio gives a measure of
the decrease in the computer time needed 
to reach a given noise variance level. 
The lattices are Wilson 
$16^{3}\times 24$, $\beta=6.0$. Note that I am using a 
\lq\lq one noise\rq\rq\, inversion method, 
meaning that a global estimate of
the operator is obtained after a single 
matrix inversion. The appropriate 
operators to apply this to are ${\bar \psi}\psi$
and local and point split ${\bar \psi}\gamma_{\mu}\psi$ 
(see Ref.~\cite{me}). Fig.~1 shows the effect
of the level of subtraction on the point-split 
charge density operator at $\kappa=0.148$. Fig.~2
shows the ratio of variances for the scalar 
(\lq\lq S\rq\rq), the four local vector (\lq\lq L VEC 1-4\rq\rq),
and the four point-split vector operators 
(\lq\lq P-S VEC 1-4\rq\rq) after 10th order subtraction, also
at $\kappa=0.148$.

\begin{figure}
\vskip 65mm
\special{illustration 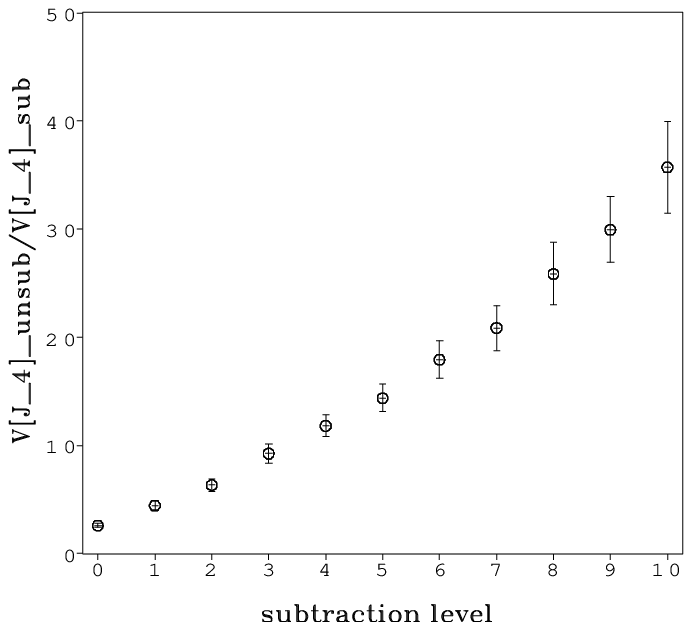}
\caption{Ratio of unsubtracted to subtracted 
noise variance for zero-momentum point-split
$J_{4}$ as a function of subtraction level 
at $\kappa=0.148$.}
\label{figure1}
\end{figure}

\begin{figure}
\vskip 65mm
\special{illustration 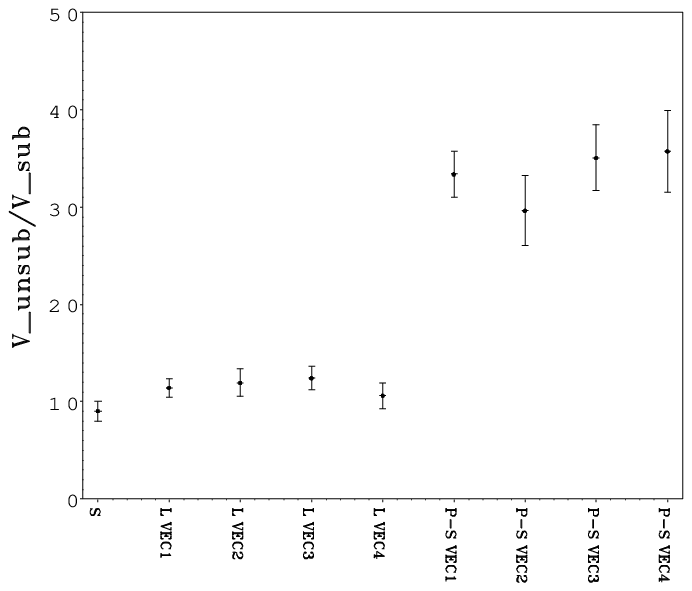}
\caption{Ratio of unsubtracted to subtracted 
variance for various operators at $\kappa=0.148$ 
after 10th order subtraction.
See the text for the meanings of the operators.}
\label{figure2}
\end{figure}

\section{Conclusions and Acknowledgments}

It has been demonstrated that a large reduction 
in the noise variance of certain lattice operators is obtained
using perturbative subtraction methods.
The method is effective for  
the scalar and local vector currents, but 
most effective for the point-split vector currents.
The method can become less effective at 
lower quark masses, depending on the operator. 
The 10th order point-split vector, 
local vector, and scalar variance ratios change 
from $\sim 35$, $\sim 12$, and $\sim 10$ at 
$\kappa =0.148$, to $\sim 25$ $\sim 10$, and $\sim 5$ at 
$\kappa =0.152$, respectively. Similar methods can be devised 
for other operators (axial, pseudoscalar, tensor) 
by implementing this algorithm in the 
context of \lq\lq 12 noise\rq\rq\, methods.

The operators condidered here are all zero momentum.
Of course for disconnected form factor
evaluations one is more interested in nonzero momentum data.
Although the results are not shown here
I have found essentially identical 
results to the above for the
momentum transformed data. These methods should be extremely 
useful in the lattice evaluations of strangeness
form factors, which are of current experimental interest.

This work is supported in part by NSF Grant No.\ 9722073 
and the National Center
for Supercomputing Applications and utilized the SGI
Origin 2000 System at the University of Illinois. The hospitality
of the University of Kentucky Physics Department, where
this work was begun, is gratefully acknowledged.


\begin{thebibliography}{9}

\bibitem{Thron} S.\ Bernardson, P.\ McCarty and C.\ Thron, 
Comp.\ Phys.\ Comm., 78 (1994) 256.
\bibitem{me} W.\ Wilcox and B.\ Lindsay, Nucl.\ Phys.\ B 
(Proc.\ Suppl.) 63A-C
(1998) 973.
\bibitem{Liu} C.\ Thron, S.\ J.\ Dong, K.\ F.\ Liu, H.\ P.\ Ying, 
Phys.\ Rev.\ D57 (1998) 1642.

\bibitem{Dong} N.\ Mathur, S-J Dong, K-F Liu and N.\
C.\ Mukhopadhyay, these proceedings.

\end{thebibliography}
\end{document}